# Effects of asymmetric contacts on single molecule conductances of $HS(CH_2)_nCOOH$ in nano-electrical junctions

Santiago Martín,[1] David Zsolt Manrique,[2] Víctor M. García-Suárez,[2] Wolfgang Haiss,[1] Simon J. Higgins,[1] Colin J. Lambert,[2] and Richard J. Nichols[*,1]

[1] Centre for Nanoscale Science and Department of Chemistry, University of Liverpool, Crown Street, L69 7ZD, U.K.

[2] Department of Physics, Lancaster University, Lancaster LA1 4YB, UK

E-mail: nichols@liv.ac.uk

# ABSTRACT

A scanning tunnelling microscope has been used to determine the conductance of single molecular wires with the configuration X-bridge-X, X-bridge-Y and Y-bridge-Y (X = thiol terminus and Y = COOH). We find that for molecular wires with mixed functional groups (X-bridge-Y) the single molecule conductance decreases with respect to the comparable symmetric molecules. These differences are confirmed by theoretical computations based on a combination of density functional theory and the non-equilibrium Green's functions formalism. This study demonstrates that the apparent contact resistance, as well as being highly sensitive to the type of the anchoring group is also strongly influenced by contact-asymmetry of the single molecular junction which in this case decreases the transmission. This highlights that contact asymmetry is a significant factor to be considered when evaluating nano-electrical junctions incorporating single molecules.

# KEYWORDS

# 1.0 Introduction

The interest in using individual organic molecules for electronic applications has prompted the development of a number of experimental[45, 44, 11, 57, 12, 34, 40, 43, 56] and theoretical methods[7, 17, 8, 47] for determining charge transport down to the single molecule level. The experimental methods include a variety of scanning probe microscopy techniques, based on either STM[57, 20, 21, 33] or conducting AFM[11, 32]. The scanning probe microscopy methods rely on forming molecular bridges between either a scanning probe tip and the sample surface[57, 20, 21, 53, 54, 9, 10] or between a probe tip and a gold nanoparticle contact.[11, 39] These techniques provide chemical attachment to the metal contacts at both ends of the molecule. Among the theoretical methods the most popular approach is based on a combination of density functional theory (DFT)[28, 25] and the non-equilibrium Green's functions formalism,[27] which facilitates the study of a large number of systems from first principles, ranging from nanotubes to atomic contacts.[11,14] The theoretical values of molecular conductance they provide for molecules sandwiched between metallic electrodes do not usually agree precisely with the experimental values but they are generally within one order of magnitude[35] and correctly reproduce the qualitative behaviours.[15]

A critical factor that determines the conductance in molecular bridges is the nature of molecule-electrode contact. Therefore, the design of effective contacts between molecule and electrodes plays a central role in molecular electronics.[5, 26, 4] One way to optimize the molecule-electrode contact is to select effective surface binding groups at the two ends of a molecule. These are typically chemisorption contacts which provide efficient electronic coupling between the electrodes and molecule. The most studied chemical contacting group used with gold electrodes is the thiol group, however carboxylate groups are one of a number of other groups which have been investigated for binding to gold electrode contacts.[10] Carboxylate or carboxylic acid groups are attractive terminal groups for molecular wires since they provide established and straightforward synthetic routes and their electrochemical adsorption behaviour is relatively well understood. Indeed, Chen et al.[10] and Martin et al.[37] have demonstrated that they can be used as effective terminal groups for molecular wires.

To date the major focus of single molecule conductance studies has been on "symmetric" contact-molecule-contact systems, where the respective chemical contacting groups at either end of the molecule are identical. However, systems where the contacting groups at either end of the molecule are different are also of potential interest for a number of reasons. These reasons may include facilitating efficient contact to differing electrodes types or directed orientation of molecules between pairs of non-similar electrodes. For instance, a polar orientation of Langmuir Blodgett films has been shown to lead to rectifying molecular junctions.[36, 2, 3, 1, 38]

The aim of this present study is to evaluate the influence of end groups and the "contact-asymmetry" of the molecular bridges on the electron-transport properties of single molecular electrical junctions. Contact asymmetry has been recognised as influencing charge transport across monolayer films, for instance monothiol self assembled monolayers (SAMs) sandwiched between a pair of electrodes, where chemical contact is only made to one electrode through the single thiol group.[29] In such cases the molecular monolayer establishes a polar orientation with respect to the contacts and consequently bias voltage polarity dependent electrical behaviour is observed including current rectification.[29] Asymmetric voltage drops at the bonded and non-bonded metal-molecule interfaces have been associated with the bias polarity dependence of the electrical behaviour.[6] The influence of contact-asymmetry has also been demonstrated for SAMs sandwiched in metal-molecule-metal junctions in cross wired electrical junctions.[30] In those studies, one terminus of the SAM featured thiol binding to the electrode while the other end employed nitro, pyridine or non-bonded contacts to the second metal electrode. It was shown that the extent of current rectification in such monolayer film junctions correlates well with the degree of electronic coupling between the chemical linker and metal electrode. Inspired by these observations of current rectification in contact-asymmetric junctions of SAMs, this present study examines experimentally and theoretically how contact-asymmetry influences the conductance of single molecules in an electrical junction. We measure and calculate the single molecule conductance of molecular wires with different combinations of end groups; either thiol groups at each end, carboxylic acids groups at each end or mixed end group systems with a thiol group at one end and a carboxylic acid group at the other.

## 2.0    Experimental Methods

The $I(s)$ method previously developed and described by Haiss *et al.*, using scanning tunnelling microscopy (STM), was used for the measurement of single-molecule conductance.[20, 21, 23, 22] In this method tunnelling current ($I$) is measured as the STM tip is withdrawn (distance, $s$) and molecular bridges are extended in the STM gap. Direct metal-to-metal contact (break junction formation) between STM tip and surface is avoided, distinguishing it from the in-situ break junction method of Tao.[57] The starting point for our measurements is the adsorption of a low coverage of molecules on a Au(111) surface prepared as has been described previously.[18] The low coverage was achieved by immersion of the gold films in 0.5 mM methanolic solutions of mercaptoalkanoic acids, alkanedithiols and α,ω-dicarboxylic acids (Aldrich, reagent grade) for 20 s. After adsorption, the samples were thoroughly washed in ethanol and blown dry with nitrogen. A Au STM tip was freshly prepared for each experiment by etching of a 0.25 mm Au wire (99.99 %) in a mixture of HCl (50 %) and ethanol (50 %) at + 2.4 V.[46]

It has been previously shown for alkanedithiol molecule bridges between gold contact gaps that there is no unique value for the single molecule conductance, with multiple single molecule conductance values being evident.[10, 33, 55, 16, 19] In the case of alkanedithiols up to three groups of peaks have been observed which has been attributed to differing contact morphologies between sulphur head groups and gold contacts.[10, 33, 19] Nevertheless, if the measurements are performed on a flat Au(111) terraced region on the sample and at low set-point current, the group of peaks with lower conductance predominate in the histogram plots of conductance values.[19] As the aim of this study is to evaluate the influence of end groups and the "contact-asymmetry" of the molecular bridges on the electron-transport properties of molecules and not the existence of multiple single molecule conductance values for these molecules we have performed all measurements on a flat Au(111) terraced and at low set-point current ($I_0$ = 1-2 nA). Under these conditions only histogram peaks corresponding to the lower conductance has been observed, see figure 1.

## 3.0 Results and Discussion

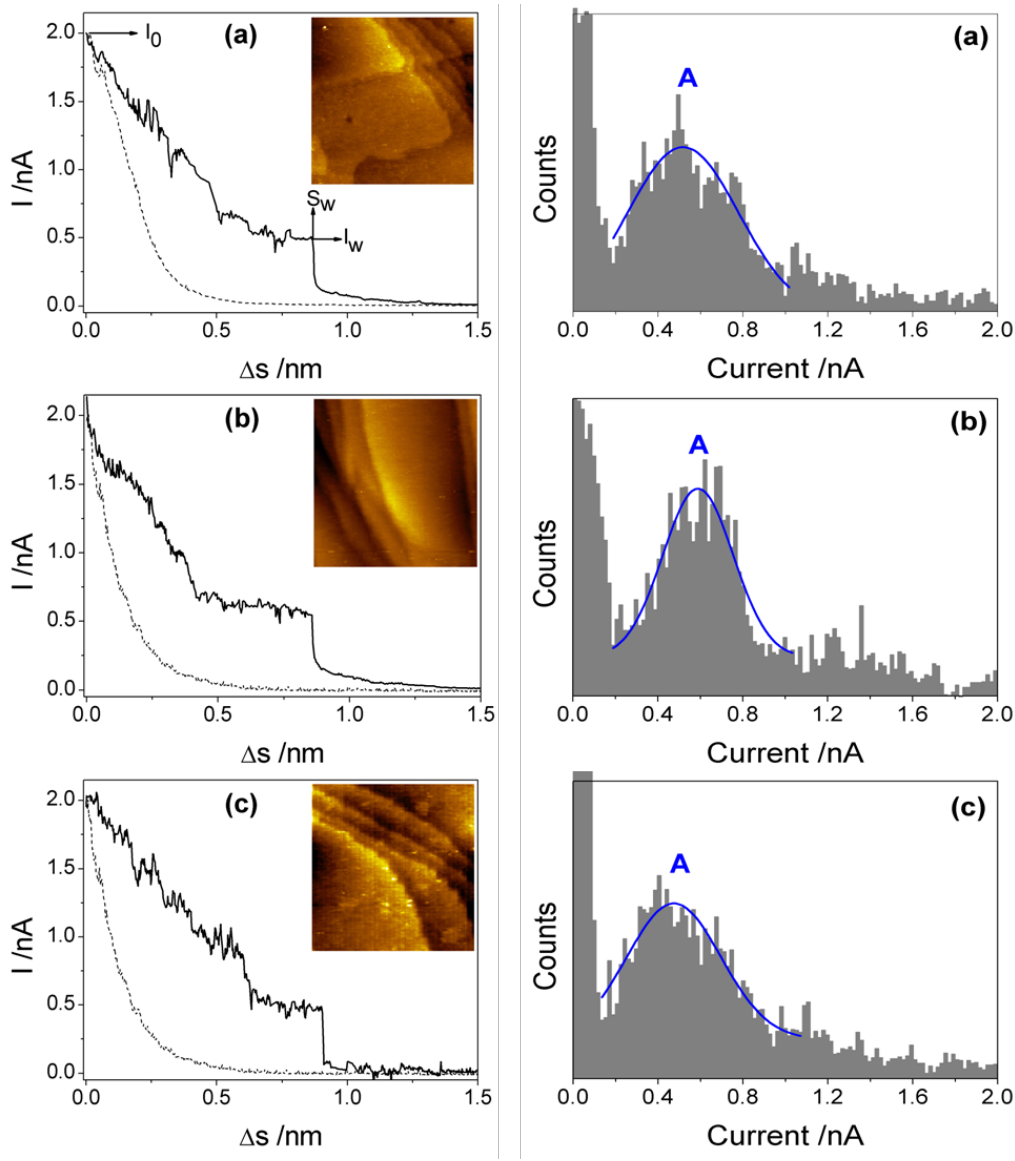

**Figure 1.** Examples of current-distance curves (left hand side) obtained as well as current histograms (right) constructed by adding together all the points of 75 individual *I*(*s*) scans exhibiting current plateaus for molecular junctions incorporating (a) HS(CH$_2$)$_7$COOH, (b) HS(CH$_2$)$_8$SH and (c) HOOC(CH$_2$)$_8$COOH. Dashed curves were recorded for bare gold substrates. The STM images (80 nm x 80 nm) presented as insets have been taken at the location where the *I*(*s*) measurements were performed. $U_t$ = 0.6 V, $I_0$ = 2 nA. Δs is the retraction distance in an *I(s)* scan from the initial set-point distance.

Figure 1a shows a typical *I*(*s*) curve in the presence and absence of 8-mercaptooctanoic acid (HS(CH$_2$)$_7$COOH) in the tunnelling gap. The decay of the current with distance is similar to that observed previously for alkanedithiols.[20] The observed current plateau of height $I_w$ (see figure 1a) is attributed to conduction through molecular wires formed between the gold tip and the substrate. 75 *I*(*s*) curves exhibiting current plateaus recorded at different locations on the substrate were analysed of approximately 1000 *I(s)* scans that were saved to yield the histograms shown in figure 1a, where a pronounced maximum marked as A can be seen. Comparing the current value obtained for HS(CH$_2$)$_7$COOH with that obtained for 1,8-octanedithiol (HS(CH$_2$)$_8$SH), figure 1b, we can observe that HS(CH$_2$)$_7$COOH has a conductance slightly smaller than that HS(CH$_2$)$_8$SH. The conductance values for HS(CH$_2$)$_7$COOH and HS(CH$_2$)$_8$SH are 0.88 ± 0.26 nS and 0.94 ± 0.22 nS, respectively. Since the polymethylene backbone of HS(CH$_2$)$_8$SH molecule is longer by one methylene unit when compared to HS(CH$_2$)$_7$COOH, these results imply a significant reduction in transmission for the asymmetric-contact system as compared to two thiol contacts (*vide infra*).

To study the influence of the metal-molecule contact on the single-molecule conductance we have also measured the single molecule conductance of alkanes with carboxylic acid terminal groups at both ends. Figure 1c shows a typical *I(s)* curve in the presence and absence of 1,10-decanedioic acid (HOOC(CH$_2$)$_8$COOH) in the tunnelling gap as well as the histogram obtained by adding together all the points of 75 *I(s)* curves exhibiting current plateaus of approximately 1000 *I(s)* scans that were saved. This exhibits a pronounced maximum that corresponds to conductance through a single HOOC(CH$_2$)$_8$COOH molecule of (0.80 ± 0.28) nS. The conductance for this molecule is only slightly smaller than that obtained for HS(CH$_2$)$_8$SH and for HS(CH$_2$)$_7$COOH in spite of having one more methylene unit than the latter compound. Since both HS(CH$_2$)$_8$SH and HOOC(CH$_2$)$_8$COOH have a C8 polymethylene chain, the slightly smaller conductance for the latter compound can be attributed to a slightly poorer electronic coupling between the carboxylate group and the gold surfaces as has been discussed previously by Chen et al. [10]. Nevertheless, it is surprising to observe that the conductance for HS(CH$_2$)$_7$COOH is only slightly larger to that obtained for HOOC(CH$_2$)$_8$COOH in spite of having one fewer methylene unit in the chain. A similar behaviour is also seen for the conductance of HS(CH$_2$)$_5$COOH ((2.5 ± 0.22) nS) when compared with HS(CH$_2$)$_6$SH and HOOC(CH$_2$)$_6$COOH ((2.6 ± 0.43) and (2.48 ± 0.38) nS, respectively). Clearer evidence of this behaviour is obtained if we compare the current histograms for molecules with the same number of methylene units (*N*=10); see figure 2. The asymmetric configuration (figure 2c, histogram on right) produces a much lower conductance for equivalent poly-methylene backbones. From these observations we conclude that alongside the influence of the nature of metal-molecule contact on the conductance, another important factor that influences the conductance of single molecule junctions is the contact-asymmetry of the molecule. In this case the presence of a thiol and carboxylic group contact on the molecular wire causes a marked decrease in transmission when compared to equivalent contact-symmetric wires (with either two thiols or two carboxylic acid groups).

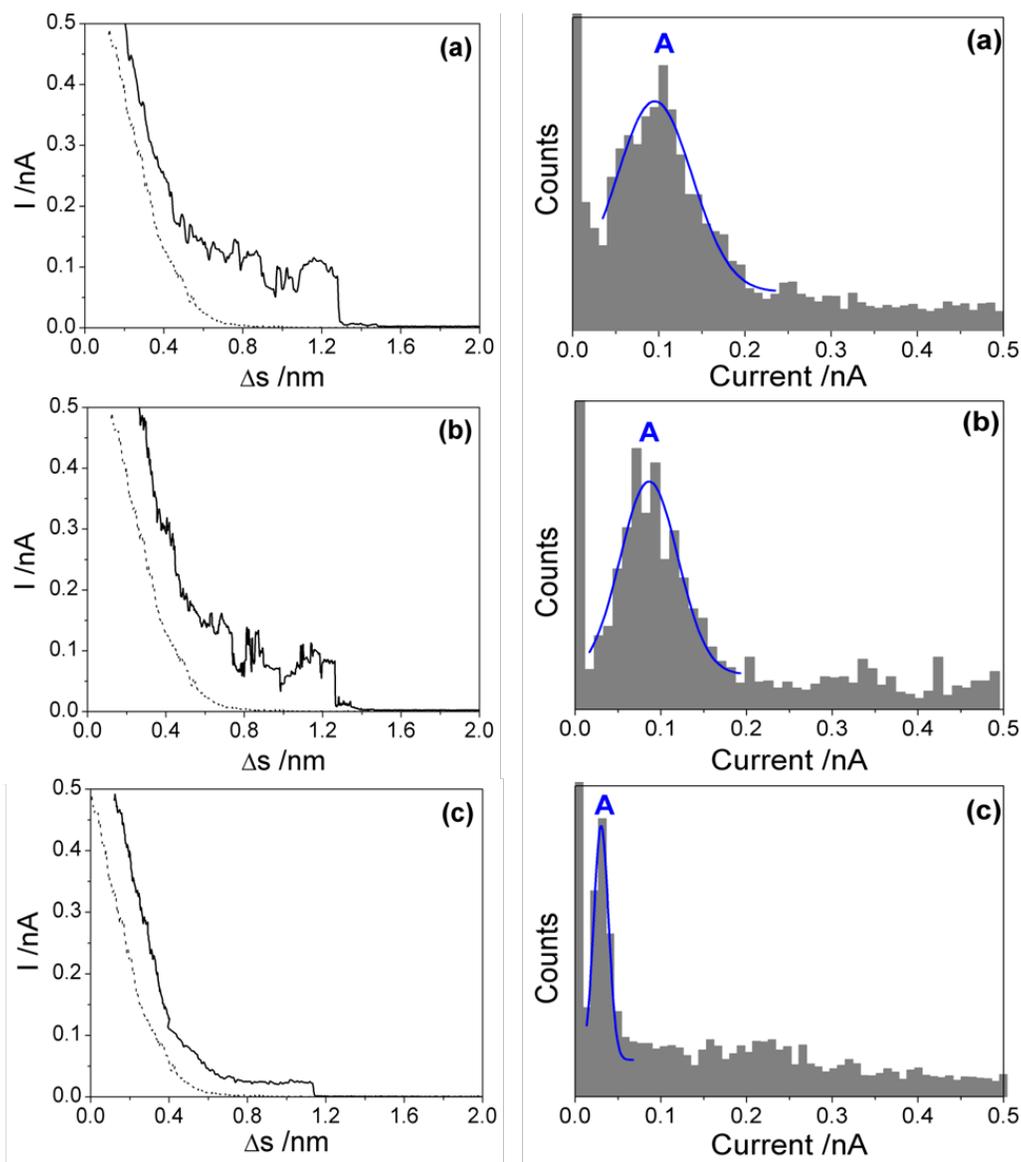

**Figure 2.** Examples of current-distance curves (left hand side) obtained as well as current histograms (right) constructed by adding together all the points of 65 individual *I(s)* scans exhibiting current plateaus for molecular junctions incorporating (a) $HS(CH_2)_{10}SH$, (b) $HOOC(CH_2)_{10}COOH$ and (c) $HS(CH_2)_{10}COOH$. Dashed curves were recorded for bare gold substrates. $U_t = 0.6$ V, $I_0 = 1$ nA. Δs is the retraction distance in an *I(s)* scan from the initial set-point distance.

Figure 3 shows the voltage dependent single molecule data for compounds $HS(CH_2)_7COOH$, $HS(CH_2)_8SH$ and $HOOC(CH_2)_8COOH$. All compounds show a similar *I(U)* behaviour (*U*=bias voltage). The *I(U)* response is close to ohmic between bias voltages of 0.8 V to -0.8 V. However, outside this voltage range the response deviates from linearity. In all cases the *I(U)* response is symmetric around zero bias even for the asymmetric molecule ($HS(CH_2)_7COOH$). This implies that the asymmetric molecule does not orient in a preferred or "polar" fashion in the gap, since such a polar orientation may be expected to yield bias dependent *I(U)* response.

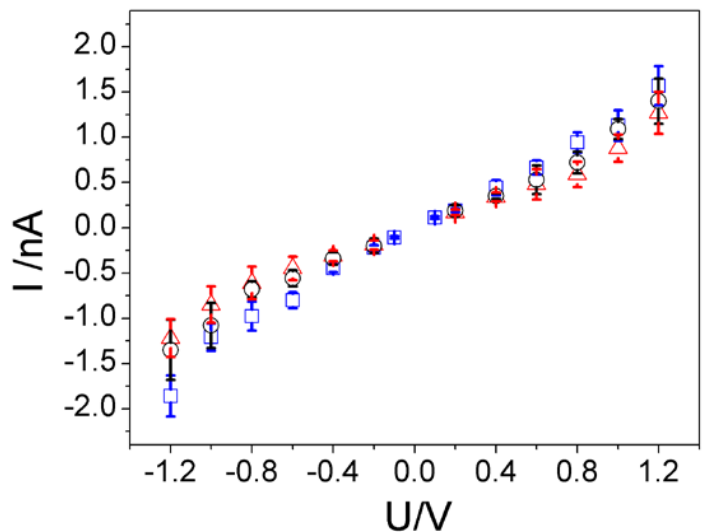

**Figure 3.** *I(U)* curves for HS(CH$_2$)$_7$COOH (black circles), HS(CH$_2$)$_8$SH (blue squares) and HOOC(CH$_2$)$_8$COOH (red triangles) from single molecule conductance. The error bars represent the standard deviation.

Kushmerick et al. have demonstrated that examining orbital topology is useful in estimating the influence of contact asymmetry on the rectifying properties of SAM junctions.[30] As explained by Kushmerick et al. the degree to which a molecular orbital provides a good conductance channel is related to its energy offset from the Fermi level and to the topology of the orbital from which the channel is derived.[30] In this respect, previous studies[24, 51, 59] have shown that orbitals which extend across the entire molecular bridge and are positioned close to the Fermi level are the most important contributors to the electronic transmission. Although the topological orbital analysis provides insight, a deeper computational analysis which quantifies the transport properties and transmission spectrum is required to more fully characterise the influence of contact-asymmetry. To achieve this, we have used the ab-initio code SMEAGOL[47], which describes both the molecule and the electrodes from first principles.

SMEAGOL employs the Hamiltonian from the density functional theory code SIESTA[49] and calculates the transport properties with the non-equilibrium Green's functions formalism. We used a double-zeta-polarized basis set to span the single-particle states and the GGA functional[42] to calculate the exchange-correlation energy. We contacted the molecule between Au electrodes and relaxed the molecular coordinates for a series of distances between the surfaces. To simulate the effect of an STM tip the sulphur atom was contacted directly to a gold atom on top of the surface (in the case of the symmetric molecule HS(CH$_2$)$_8$SH both sulphur atoms were contacted to a gold adatom to maintain the symmetry of the system), which was also relaxed. The initial distances between the surfaces corresponded to roughly the length of the relaxed molecule without leads plus 2 Å on both sides plus the gold adatoms, which resulted in 19.90 Å, 16.85 Å and 17.12 Å for HS(CH$_2$)$_8$SH, HS(CH$_2$)$_7$COOH and HOOC(CH$_2$)$_8$COOH junctions, respectively. For these distances the molecules were slightly stretched since their energy was not minimized and the oxygens of the carboxylate contact were in the on-top surface position, as opposed to the more stable bridge position that appeared when the molecule was not stretched. We studied the conductance as a function of the separation between the leads for a range of distances to take into account different configurations. Some typical cases for the HS(CH$_2$)$_7$COOH molecule are shown in figure 4 and their conductances, along with the conductances of the other two molecules, are listed in table 1. Notice that the evolution with distance is not the same for each molecule due to their different lengths and contact groups. The configurations which gave results similar to experiments corresponded to the initial distances, excluding the HOOC(CH$_2$)$_8$COOH which was slightly more stretched (18.66 Å). The resulting conductances are 1.8x10$^{-4}$ $G_0$ (13.9 nS) for HOOC(CH$_2$)$_8$COOH, 2.2x10$^{-4}$ $G_0$ (18.5 nS) for HS(CH$_2$)$_7$COOH, and 2.2x10$^{-4}$ $G_0$ (18.5 nS) for HS(CH$_2$)$_8$SH, where $G_0 = 2e^2/h \approx 77.4$ µS. These values are at least one order of magnitude higher than experimentally measured values, consistent with previous DFT computations which have overestimated conductance values.[35] These cases are shown in figure 5, which clearly indicates the transport is in the tunnelling regime since $E_F$ is in the HOMO-LUMO gap. We also show the asymmetric molecule with one alkane unit more (HS(CH$_2$)$_8$COOH, corresponding to a distance of 18.56 Å) to demonstrate that the asymmetric configuration produces a much lower conductance for equivalent poly-methylene backbones. In other words for the (CH$_2$)$_8$ series the theoretically computed conductance (at $E_F$) follows the trend:

$$HS(CH_2)_8SH \approx HOOC(CH_2)_8COOH \quad > \quad HS(CH_2)_8COOH$$

This agrees with the experimental observation of the negative impact of the contact asymmetry on conductance for this system. Also notice the main difference between HS(CH$_2$)$_8$COOH and HS(CH$_2$)$_7$COOH appears in the gap, whereas the HOMO and LUMO features are similar in both cases. The shapes of the transmission curves are smooth and very similar for all molecules in the vicinity of $E_F$, and the main differences appear around $E_F = -1.5$ to -2.5 eV and $E_F = +2$ eV, with features which are very sensitive to the chemical composition of the molecular bridges and head group combinations.

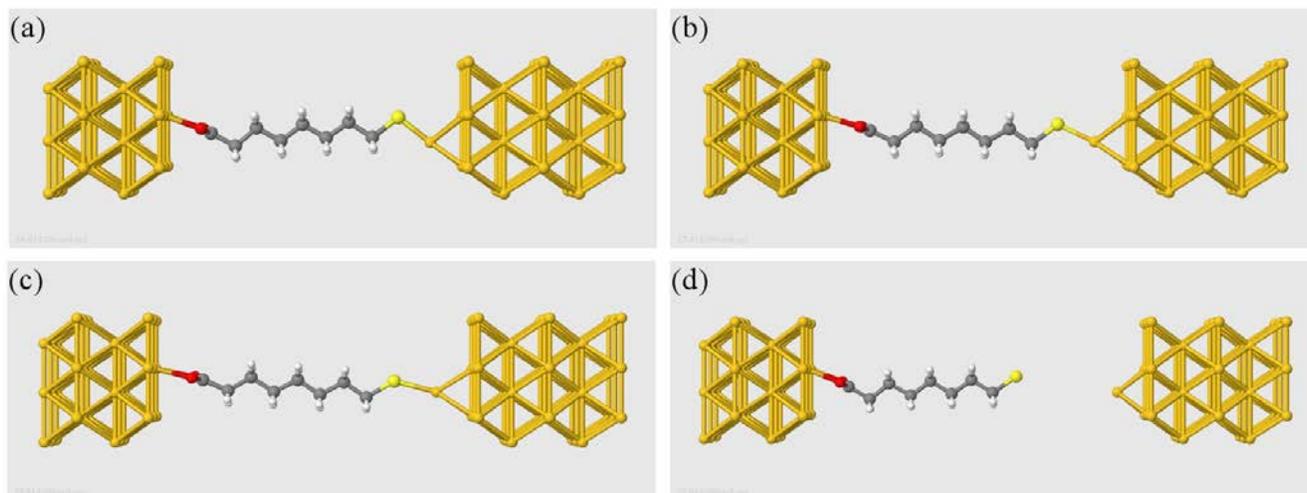

**Figure 4.** Some of the configurations of the HS(CH$_2$)$_7$COOH molecule used to calculate the conductance with SMEAGOL. The distances between electrodes are (a) 16.61 Å, (b) 17.41 Å, (c) 18.81 and (d) 21.01 Å.

| | | | | | |
|---|---|---|---|---|---|
| HS(CH$_2$)$_7$COOH | 16.41 Å | 16.81 Å | **17.21 Å** | 17.61 Å | 18.01 Å |
| *G* (nS) | 76.4 nS | 16.6 nS | 18.5 nS | 24.8 nS | 26.4 nS |
| HS(CH$_2$)$_8$SH | 19.41 Å | 19.81 Å | **20.21 Å** | 20.61 Å | 21.01 Å |
| *G (nS)* | 22.0 nS | 22.8 nS | 18.5 nS | 19.8 nS | 19.2 nS |
| HOOC(CH$_2$)$_8$COOH | 17.86 Å | 18.26 Å | **18.66 Å** | 19.06 Å | 19.46 Å |
| *G* (nS) | 4.6 nS | 9.3 nS | 13.9 nS | 8.5 nS | 0.2 nS |

**Table 1.** Conductance values of some typical molecular configurations around the initial separation. Data is tabulated for 5 different separations between electrodes and 3 molecules are shown for comparison.

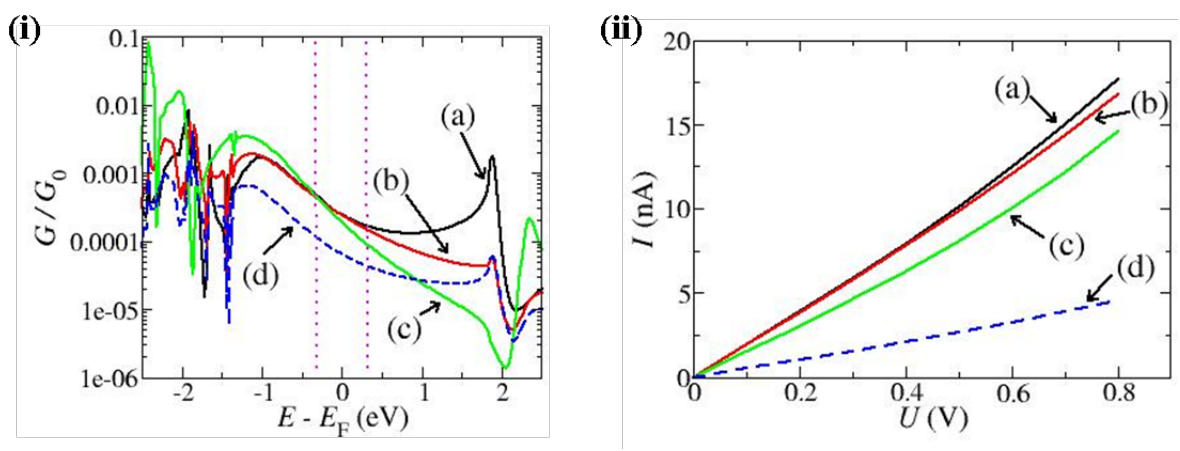

**Figure 5(i).** Transmission coefficients of HS(CH$_2$)$_8$SH (black line, marked (a)), HS(CH$_2$)$_7$COOH (red line, (b)), HOOC(CH$_2$)$_8$COOH (green line, (c)) and HS(CH$_2$)$_8$COOH (blue dashed line, (d)), sandwiched between Au electrodes. Transmission at $E_F$ is much less for HS(CH$_2$)$_8$COOH as a result of contact asymmetry. The vertical dotted lines (purple) represent a bias window of 0.6 V, for comparison with experiments. Figure 5(ii) The calculated *I(U)* curves for bias voltages up to 0.6 V.

For shorter separations, which minimize the energy, the molecules are compressed and somewhat deformed. For larger separations the bond between the sulphur and the surface approaches cleavage and the conductance rises again until the bond definitively breaks and the conductance decreases exponentially. In the case of HOOC(CH$_2$)$_8$COOH the bond between the oxygens and the surface breaks more abruptly and it is more difficult to observe the rise in conductance, as can be seen in Table 1. The best agreement is found for distances slightly larger than the molecular length. This is consistent with the notion that the molecules are put under tension as they are fully extended and stretched in between the contacts in the *I*(*s*) or break junction experiments. The force required to cleave a molecular junction in a scanning probe microscopy break junction experiment has been determined as (1.5 ± 0.2) nN.[58] This corresponds to the force needed to break a Au-Au bond.[48] This indicates that relatively large forces are applied to the molecules as they are stretched to cleavage in the single molecule conductance experiments.

To understand the results of figure 5, we now develop a heuristic tight-binding model,[41] which captures the essential quantum mechanics of superexchange between conductance electrons of ideal 1-D gold leads and two degenerate HOMO orbitals located on both ends of the molecule associated to either the S or COO groups. The model is illustrated in figure 6 and comprises two energy levels $\varepsilon_1$ and $\varepsilon_2$, coupled by an intra-molecular hopping element *t* and respectively coupled to left and right leads by hopping elements $g_L$ and $g_R$.

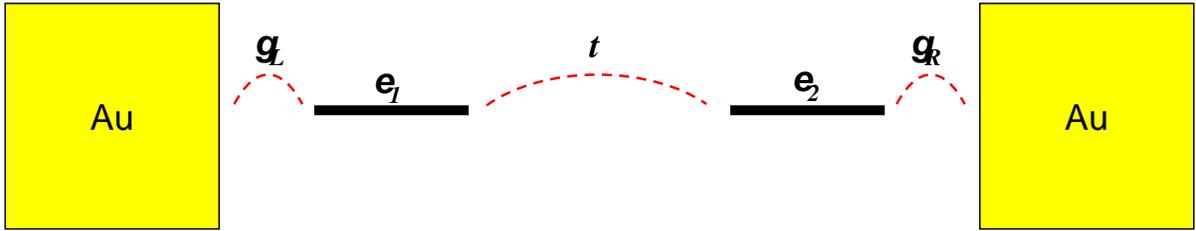

**Figure 6.** Schematic representation of the simple model used to reproduce the ab-initio transmission coefficients around the HOMO orbital.

By solving the Schrödinger equation for this structure,[31] we find the following equation for the electron transmission coefficient:

$$T(E) = \frac{4\Gamma_L \Gamma_R t^2}{[(E-\varepsilon_1)(E-\varepsilon_2) - \Gamma_L \Gamma_R - t^2]^2 + [(E-\varepsilon_1)\Gamma_R + (E-\varepsilon_2)\Gamma_L]^2},$$

where $\Gamma_{L,R} = g_{L,R}^2 \rho_{L,R}$ and $\rho_{L,R}$ is the density of states in the leads. To gain insight into ab initio results, we have varied the parameters in the above model, to reproduce the results of figure 5. The results are shown in figure 7. The width of each resonance depends on the couplings to the leads ($\Gamma_L, \Gamma_R$), the position of the energy levels ($\varepsilon_1, \varepsilon_2$) and the magnitude of the intra-molecular coupling (*t*). The shape and position of these curves can therefore be tuned to make them almost coincide with the curves of figure 5 (excluding of course features above the Fermi level associated to the LUMO). Figure 7a shows results for $\varepsilon_1 = \varepsilon_2$ and $\Gamma_L, \Gamma_R = 0.6$, while Figure 7b shows small changes when $\Gamma_L, \Gamma_R = 0.4$. Figure 7c, on the other hand, shows the situation for different coupling at the right and left leads ($\Gamma_L \neq \Gamma_R$), which represents the asymmetric case. This asymmetric case yield a line which falls between the symmetric results; indeed where $\varepsilon_1 = \varepsilon_2$, most reasonable choices of the parameters $\Gamma_L, \Gamma_R, t$ yields curves for the asymmetric case which fall between the symmetric ones. The only way the curve for the asymmetric cases ($\Gamma_L \neq \Gamma_R$) can be made to lie significantly below the other two is if the intra-molecular coupling ($t$) is smaller than that for the symmetric cases; see Figure 7d, where $t$ has been reduced from 0.015 to 0.008 . This behaviour can arise if in the asymmetric case the molecule is more stretched or the matrix element between $\varepsilon_1$ and $\varepsilon_2$ is otherwise reduced due to the coupling of the alkane chain to different atoms (S and COO). We also note that the similarity between these and the ab initio results suggests that electron transmission near the Fermi energy is dominated by superexchange with the double-degenerate HOMO levels.

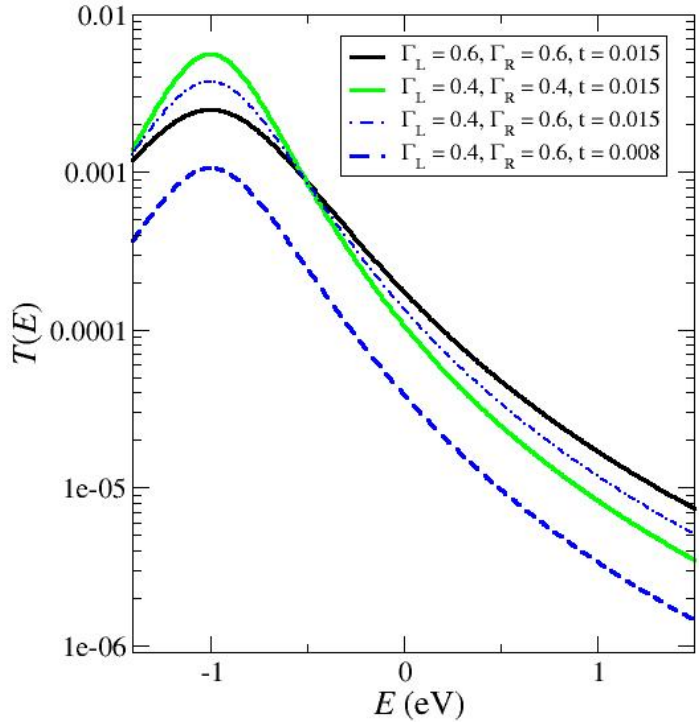

**Figure 7.** Transmission coefficients obtained by using the simple model described in figure 6. We used $\varepsilon_1 = \varepsilon_2 = -1$ (in atomic units) and varied the coupling strengths to the leads ($\Gamma$) and the inter-level coupling $t$ which represents the alkane chain.

To further explore the gap separation at which the molecular junction is cleaved careful calibration of the tip-to-substrate distance is required. As explained in the literature,[23] this calibration is achieved by recording tunnelling current ($I$) versus distance curves ($s$) and obtaining $d\ln(I)/ds$ in the distance range of interest and extrapolating to the point contact conductance to obtain the zero tip to substrate reference separation. The retraction distance measured in an $I(s)$ scan can then be recalibrated to the separation between tip and sample. The distance at which current steps in the $I(s)$ curves occur ($s_w$ in figure 1a) plus the initial set-point distance ($s_0$) can be compared to the length of the molecule. The break-off distance ($s_{break-off}$) is then $s_0+s_w$. These break-off distances are shown in Table 2. The experimentally measured break-off distances are very close to values computed by molecular modelling for each of the molecules fully extended between gold contacts, with the distance being that between the respective gold contact atoms at each end of the molecular wire.

| Molecules | $S_{calc}$ (Å) (all-*trans* conformation) | $S_{break\text{-}off}$ (Å) (break off point) | $d_{Au\text{-}Au}$ (Å) |
|---|---|---|---|
| HS(CH$_2$)$_7$COOH | 13.9 | 14.5 ± 1.8 | 15.3 |
| HS(CH$_2$)$_8$SH | 14.0 | 13.3 ± 2.1 | 16.4 |
| HOOC(CH$_2$)$_8$COOH | 14.9 | 15.6 ± 1.6 | 18.7 |

**Table 2.** A listing of the measured break-off point ($s_{break\text{-}off}$) using the *I*(*s*) method for three different molecules and the molecular length obtained for these molecules assuming all-*trans* conformation ($s_{calc}$) and the distance between the Au electrodes ($d_{Au\text{-}Au}$) for the theoretically computed molecular configuration which give conductance values which best agree with the experimental results.

The molecular conductance of these molecules is now analysed as a function of molecular length by varying the number of -CH$_2$- units in the polymethylene (alkane) chain (*N*). Figure 8 shows the dependence of the logarithm of $\sigma_M$ on the number of –CH$_2$- groups in the molecule. The linear fits indicate that the data sets can be described by $G = G´·\exp(-\beta_N N)$, which suggests superexchange as the conduction mechanism in all molecules.[14] In this equation, *G* is the conductance; *N* the number of methylene units; *G´* is a constant determined by the molecule-electrode coupling which reflects the terminal groups contact conductance; and $\beta_N$ is the tunnelling decay constant, an important parameter that describes the efficiency of charge transport by tunnelling through the molecules. All these molecules have a large HOMO-LUMO gap and therefore the $\beta_N$ value is expected to be rather invariant with the number of methylene units in the alkyl chain.[13, 14] The tunnelling decay constant, $\beta_N$, for alkanedithiols and mercaptoalkanoic acids is the same within experimental error and in the expected range (0.89 ± 0.03 and 0.87 ± 0.04, respectively). Nevertheless, the $\beta_N$ value for alkane dicarboxylic acids series is somewhat less than for the other molecular series, 0.78 ± 0.07 per CH$_2$ unit. However, the most noteworthy feature is that the line describing the asymmetric molecular junctions (HOOC-(CH$_2$)$_N$-SH) lies significantly lower than the symmetric ones.

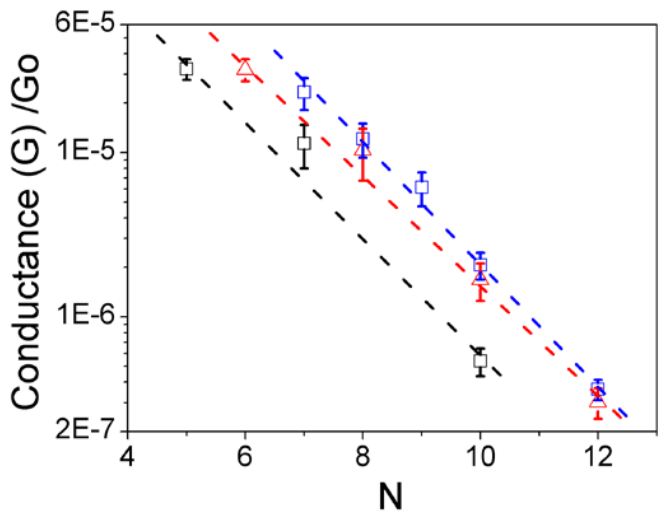

**Figure 8.** Semilog plot of conductance versus molecular length for α,ω-mercaptoalkanoic acids (black circles), α,ω-alkanedithiols (blue squares) and α,ω-dicarboxylic acids (red triangles). The dashed lines are the linear fits to the data. The error bars represent the standard deviation. $U_t = 0.6$ V.

Extending the fit in Figure 8 to the origin, we can obtain the value of the apparent contact resistance ($1/G´$) for these molecular series. If we compare the effective contact resistance for the alkane dicarboxylic acids (250 $G_0^{-1}$ or 3.2 MΩ; where $G_0 = 2e^2/h = 77.4$ μS) with the value obtained for the alkanedithiol series (67 $G_0^{-1}$ or 0.87 MΩ) it is about four times higher. $1/G´$ is an aggregate contact resistance that includes the resistance of both the top and bottom contacts. Consistent with earlier studies the difference in the *A* values arises from the different transmission properties of the Au-S-alkane and Au-OOC-alkane contacts.[52, 50] For the mercaptoalkanoic acids series we observe that the apparent contact resistance is 333 $G_0^{-1}$ (4.3 MΩ), which is much higher than that obtained for alkanedithiols and also higher than that obtained for alkane dicarboxylic acids in spite of having one Au-S-alkane contact. Following the foregoing discussion, this can be qualitatively understood as arising from the contact-asymmetry which leads to an overall reduction in transmission through the molecular junction.

## 4.0    Conclusions

In conclusion, we have demonstrated using the *I(s)* method that the efficiency of the charge transport through single molecular bridges formed between the gold STM tip and the substrate is influenced not only by the nature of metal-molecule contact but by the contact asymmetry of the molecular junction as well. The experimentally determined single molecule conductance of $HS(CH_2)_7COOH$ is only slightly smaller than that obtained for $HS(CH_2)_8SH$ and slightly larger than that obtained for $HOOC(CH_2)_8COOH$ in spite of having one less methylene unit in the alkyl chain. Likewise, in the series $HS(CH_2)_{10}SH$, $HOOC(CH_2)_{10}COOH$, $HS(CH_2)_{10}COOH$, which have equal length polymethylene chains ($-(CH_2)_{10}-$), the asymmetric configuration produces the lowest conductance. This adverse influence of contact-asymmetry is also confirmed by theoretical computations based on a combination of density functional theory and the non-equilibrium Green's functions formalism, where it has been found that the conductance of $HS(CH_2)_7COOH$ is identical to $HS(CH_2)_8SH$ despite having one $CH_2$ group less. We also note that the computed values are closest to experimental values when the molecules are slightly stretched in the theoretical calculations, rather than being completely relaxed into non-fully extended configurations within the junction. We have also studied the molecular conductance of these molecular series as a function of molecular length. We obtained similar $β_N$ values for alkanedithiols and mercaptoalkanoic acids and slightly smaller values for alkane dicarboxylic acids. Finally, we have observed that the apparent contact resistance ($1/G´$) as well as being highly sensitive to the type of the anchoring group is also influenced by contact-asymmetry of the molecular junction, with the asymmetric $HS(CH_2)_7COOH$ junctions showing the highest $1/G´$ values.

ACKNOWLEDGMENT

We thank the NWGrid for computing resources. This work was supported by EPSRC under grant EP/C00678X/1 (Mechanisms of Single Molecule Conductance), the European Commission (for a Transfer of Knowledge Marie-Curie project, contract number MTKD-CT-2005-029864), Qinetiq, and the British Department of Trade and Industry, Royal Society and Northwest Regional Development Agency. S. M. acknowledges a postdoctoral fellowship from Ministerio de Educacion y Ciencia of Spain.